# Boron: a Hunt for Superhard Polymorphs

A. R. Oganov[a,b] and V. L. Solozhenko[c]

[a]Department of Geosciences, Department of Physics and Astronomy, and New York Center for Computational Sciences, Stony Brook University, Stony Brook, New York 11794-2100, USA
[b]Geology Department, Moscow State University, 119992 Moscow, Russia
[c]LPMTM-CNRS, Université Paris Nord, 93430 Villetaneuse, France

Abstract—Boron is a unique element, being the only element, all known polymorphs[1] of which are superhard, and all of its crystal structures are distinct from any other element. The electron-deficient bonding in boron explains its remarkable sensitivity to even small concentrations of impurity atoms and allows boron to form peculiar chemical compounds with very different elements. These complications made the study of boron a great challenge, creating also a unique and instructive chapter in the history of science. Strange though it may sound, the discovery of boron in 1808 was ambiguous, with pure boron polymorphs established only starting from the 1950s–1970s, and only in 2007 was the stable phase at ambient conditions determined. The history of boron research from its discovery to the latest findings pertaining to the phase diagram of this element, the structure and stability of β-boron, and establishment of a new high-pressure polymorph, γ-boron, is reviewed.

An element with a huge range of applications from nuclear reactors to superhard, thermoelectric and high-energy materials, boron is also arguably the most complex element in the Periodic Table. The history of boron research is full of disputes, with mistakes made even (or mainly?) by great scientists. At times this history may even be read like a detective story. A story that we briefly recount here, focussing on the most recent discovery – that of a high-pressure superhard phase of boron called γ-$B_{28}$ [1].

A boron-containing mineral borax, $Na_2[B_4O_5(OH)_4]\cdot 8H_2O$, has been known since ancient times and its name derives from Arabic "buraq", which means "white". In 1702, starting from borax, Wilhelm Homberg obtained a snow-white powder that he called "sedative salt", now known as metaboric acid, $HBO_2$. The next stage, marked by the "double discovery" of this element, was at the time of scientific rivalry between great English (Humphry Davy) and French (Louis Joseph Gay-Lussac and Louis Jacques Thenard) chemists. On June 21, 1808 Gay-Lussac and Thenard announced the discovery of the new element, which they called "bore" (the element is still called this name in French). They obtained boron by reduction of boric acid with potassium[2] [2].

---

[1]Rhombohedral α-$B_{12}$ and β-$B_{106}$ phases (with 12 and ~106 atoms in the unit cell, respectively), tetragonal T-192 (with 190-192 atoms/cell) and orthorhombic γ-$B_{28}$ (with 28 atoms in the unit cell).
[2] The story of how Gay-Lussac and Thénard obtained potassium is also quite instructive. Davy's discovery of this element by electrolysis caused a great excitement among chemists. Emperor Napoleon I, who awarded a prestigious prize to English chemist Humphry Davy, wished to foster similar kind of research (discovery of new elements by electrolysis) and presented Davy's competitors, Gay-Lussac and Thénard, with a very large electric battery.

Within days, on June 30, 1808, Humphrey Davy submitted to the Royal Society of London an article on the discovery of a new element (which he called boracium)[3] [3]. Faithful to his style, which has led to the discovery of a whole pleiad of elements, Davy prepared boron by electrolysis. The first detective twist, apart from the remarkably close dates of the two independent discoveries, is that both discoveries did not produce a pure element. It is now clear that both groups synthesized compounds containing no more than 50% of boron [4]. Should the works of Gay-Lussac, Thenard and Davy be still considered as discoveries? It is hard to answer positively, but if we answer negatively, then it will be exceedingly hard to say who actually discovered the element, as we discuss below.

One of the prominent scientists, who proved that Gay-Lussac, Thenard and Davy did not deal with a pure element, was Henri Moissan. In 1895 he prepared the element by reduction of $B_2O_3$ with magnesium in a thermite-type reaction [4]. However, even Moissan's material was far from being a pure element. It is often quoted that 99%-pure boron was synthesized by E. Weintraub [5] in 1911, and while his methods were certainly advanced compared to the previous works, there are reasons for doubt, as pure boron polymorphs are documented only after 1957.

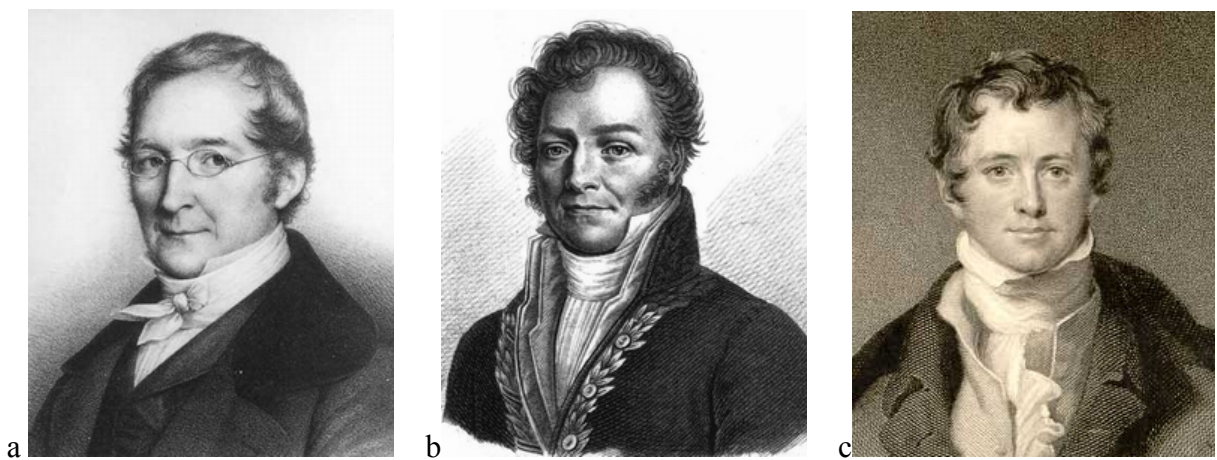

Fig. 1. Discoverers of boron: (a) Joseph Louis Gay-Lussac (1778-1850), (b) Louis Jacques Thenard (1777-1857), (c) Humphry Davy (1778-1829).

After the element itself got more or less established, a race for discovery of boron polymorphs slowly began. And that race was equally complex and full of misdiscoveries. Already in 1857 Friedrich Wöhler and Henri Sainte-Claire Deville [6, 7] heating up boron oxide and aluminum obtained three forms of boron. On the basis of hardness and luster, they drew an analogy with carbon polymorphs and called these forms diamond-like, graphite-like, and charcoal-like (amorphous). Amorphous form had the same properties as the material synthesized by Gay-Lussac and Thenard (which, as we now know, was not pure boron), while the "diamond-like" and "graphite-like" forms were later proven to be compounds containing no more than 70% of boron [4]. Thus, many, if not most, of the great scientists who studied boron, fell victims of this element's extreme sensitivity to even small amounts of impurities. This sensitivity is evidenced

---

Disappointingly, the battery turned out to be not nearly as powerful as was expected. Gay-Lussac and Thénard, however, managed to prepare potassium by heating up potash and iron.

[3] The material obtained by Davy appears to have been metallic, whereas pure boron phases are all semiconducting.

by the existence of such very boron-rich compounds (with unique icosahedral structures) as $YB_{65.9}$, $B_6O$, $NaB_{15}$, $B_{12}P_2$, $B_{13}P_2$, $B_{13}C_2$, $MgAlB_{14}$, $AlC_4B_{40}$, $NiB_{50(?)}$, $B_{50}C_2$, $B_{50}N_2$, $PuB_{100(?)}$ (e.g., [8]). In particular, the structure of $YB_{66}$ is an icon of structural complexity – it contains 1584 atoms in the unit cell [9].

It is fair to state that boron still remains a poorly understood element. At least 16 crystalline polymorphs have been reported [8], but crystal structures were determined only for 4 modifications and most of the reported phases are likely to be boron-rich borides rather than pure elemental boron [8, 10, 11]. Until 2007, it was the only light element, for which the ground state was not known even at ambient conditions. And none of the polymorphs reported before 1957 actually correspond to pure boron. Most of the discoveries related to pure boron discoveries were done in two "waves", i.e. in 1957–1965 and 2001–2009.

The first wave was led by researchers from the Cornell University and General Electric (GE) Corporation. The so-called I-tetragonal phase (or T-50, because it contains 50 atoms in the unit cell), produced in 1943 jointly at Cornell and GE [12], was the first one, for which the structure was solved – in 1951 [13, 14]. This structure appeared in many books, notably in Pauling's seminal *The Nature of the Chemical Bond* [15], where it was the only boron structure depicted[4]. However, this "well-established" phase was proven to be a compound [10, 11, 16] of composition $B_{50}C_2$ or $B_{50}N_2$.

The first pure boron phase discovered was $\beta$-$B_{106}$ [17], the structure of which turned out to be extremely complex and was solved only several years later [18]. This discovery was shortly followed by the discoveries of $\alpha$-$B_{12}$ phase at GE in 1958 [19] and T-192 phase at Polytechnic Institute of Brooklyn in 1960 [20] (the structure of the latter was so complex that it was solved only in 1979 [21]). All these structures contain $B_{12}$ icosahedra and are shown in Fig. 2.

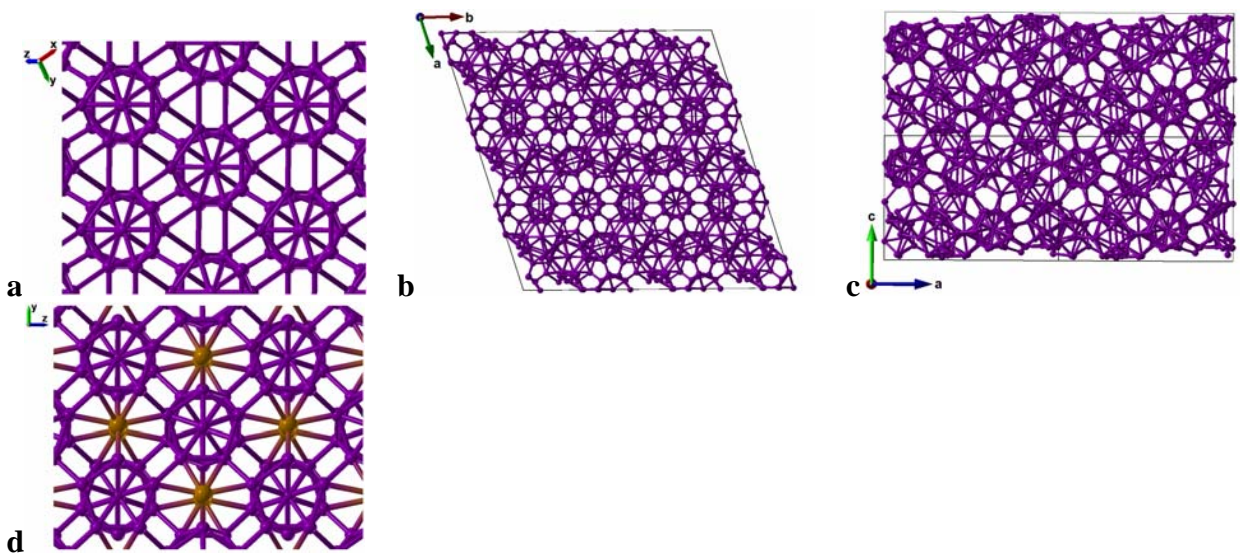

---

[4]Although Pauling used a "wrong" phase for illustrating chemical bonding in boron, the ideas themselves were largely correct.

Fig. 2. Crystal structures of boron polymorphs: (a) α-$B_{12}$, (b) β-$B_{106}$, (c) T-192, (d) γ-$B_{28}$. Panels (a) and (d) are from [1].

Considering the wealth of different "boron polymorphs" reported in the literature, Amberger and Ploog [10] even suggested that only two known phases correspond to pure boron, namely, α-$B_{12}$ and T-192, and possibly β-$B_{106}$. They obtained boron by CVD using a mixture of $BBr_3$ and $H_2$ at temperatures of 1200–1600 K, by deposition on Ta wires in the absence of any foreign atoms. This way they observed amorphous boron, α-$B_{12}$ and T-192 phases, and occasionally β-$B_{106}$. T-50 phase (obtained earlier by a similar method [12]) was never synthesized. Because of the sensitivity of boron to impurities different samples of the same polymorph show important differences in structural and, as a result, in thermodynamic properties. The relative stability of boron phases is still experimentally unresolved even at ambient conditions [22]. It was, for instance, a matter of a debate (until 2007) whether α-$B_{12}$ or disordered β-$B_{106}$ is stable at ambient conditions.

As we mentioned, much of the progress was done at GE. At that time, GE amassed a unique group of researchers with the aim of enabling industrial-scale synthesis of diamond. Furthermore, GE researchers synthesized cubic BN, an advanced substitute for diamond in cutting and abrasive tools. Both synthetic diamond and cubic BN (commercialized under the name "borazon") turned into multimillion-dollar industries. GE was interested in boron, because of its extreme hardness and because of its highly tunable electrical conductivity. At least since the works of Sainte-Claire Deville and Wöhler in 1850s, boron was known to be the second hardest element after carbon (viz. diamond), and Weintraub [5] even entertained ideas that under certain fabrication protocol boron could become harder than black diamond (variety of diamond called "carbonado")[5].

In 1964, Robert Wentorf of GE, one of the pioneers of high-pressure synthesis of materials and the main author of the synthesis of cubic BN, turned to study the behavior of boron under pressure. At pressures above 10 GPa and temperatures of 1800-2300 K, he found that both β-$B_{106}$ and amorphous boron transformed into another, hitherto unknown, phase [23]. Wentorf reported a qualitative diffraction pattern of the new material and described the changes of the density and electrical conductivity across the phase transition. For that time, it was a state-of-the-art work, but nevertheless it was not accepted by the community from the beginning. In a period of 1965–2008 Wentorf's paper was cited only 6 times (in spite of being published in a very prestigious journal, *Science*) and only by papers dealing with boron indirectly. Classical papers (e.g., the paper of Amberger and Ploog [10] on pure boron phases) never even mention Wentorf's paper! Furthermore, Wentorf's diffraction data were deleted from Powder Diffraction Files Database. One can speculate about the reasons for such a distrust, but almost certainly these were (1) absence of chemical analysis in Wentorf's paper (neither for the starting material, nor for the product – and this in principle disqualifies any experimental work on boron, as we saw above), (2) lack of crystal structure determination (there was a lasting doubt that Wentorf's material was a mixture of phases – which, indeed, it most likely was).

---

[5] E.g., Weintraub wrote that "(pieces of boron) are very hard and scratch with ease the known hard substances except diamond", "in further continuation of the work additional toughness may be imparted to boron and the product becomes a cheap substitute for black diamond" and "Will it be possible to approach the properties of diamond or perhaps by combining boron and carbon even exceed diamond in its hardness? I can only say that we are working on this problem." [5]

The second wave of boron studies was probably catalyzed by the 2001 unexpected discovery of superconductivity in $MgB_2$ [24]. It was clear from the beginning that superconductivity of this compound is due to the graphite-like sublattice of boron atoms (e.g., [25]), and very soon elemental boron was subjected to high pressures in search for superconductivity. In 2001, compressing $\beta$-$B_{106}$ at room temperature, Eremets *et al*. [26] indeed observed metallization at 160 GPa, and the metallic state displayed superconductivity (with the value of $T_c$ reaching 11.2 K at 250 GPa). The structure of this metallic phase was not determined, but subsequent experiments [27] suggested that room-temperature compression of $\beta$-$B_{106}$ results in pressure-induced amorphization at 100 GPa. This implies that there is a kinetically hindered phase transition to some unknown crystalline phase below 100 GPa. The likely problem of room-temperature experiments is metastability. Using laser heating to overcome kinetic barriers, Ma *et al*. [28] have found that $\beta$-$B_{106}$ transforms into the T-192 phase above 10 GPa at 2280 K. This has proven that the T-192 phase is not only a pure boron phase, but also has a stability field at high pressures and temperatures. Its stability field was further constrained in [1].

At the same time, the stable phase at ambient conditions remained unknown, putting XX and XXI century chemists in shame. The debate whether $\alpha$-$B_{12}$ or $\beta$-$B_{106}$ is stable at ambient conditions, was finally resolved in 2007-2009 by *ab initio* calculations of three different groups [29–31], which used different approaches, but all concluded in favor of $\beta$-$B_{106}$ and against common intuition that favored the much simpler $\alpha$-$B_{12}$ structure.

Another major result came up from Chen (experiments done in February 2004 at Stony Brook University) and Solozhenko (experiments done in April 2004 at the Université Paris Nord and HASYLAB-DESY, and in July 2004 at the Bayerisches Geoinstitut). Both groups found a new phase of boron at pressures above 10-12 GPa and temperatures above 1500 K. Further evolution of ideas and events completed the "second wave of boron research", resolved many old problems and by itself could deserve a detective novel.

Although Chen and collaborators managed to determine the unit cell parameters of the new phase, neither group succeeded in solving its structure, in spite of intense research and repeated experiments during several years. In 2006 Chen posed this problem to Oganov, whose method for predicting crystal structures [32] could be used for solving this problem. The structure was solved within one day[6], and its simulated diffraction pattern coincided with the experimental one (to make the test challenging and unbiased, Oganov and his team of theoreticians did not have access to experimental diffraction data and thus, the comparison was done in a "blind" way). The structure, thus confirmed by experiment, was indeed unique – it is a NaCl-type arrangement of two types of clusters, $B_{12}$ icosahedra and $B_2$ pairs[7] [1]. Detailed investigations showed that these two clusters have very different electronic properties (Fig. 3) and there is a charge transfer (of ~0.5 e) from $B_2$ to $B_{12}$ [1], and this is correlated with the strong IR absorption and high dynamical charges on atoms. This partially ionic phase was named $\gamma$-$B_{28}$ [1]. Our measurements [33] showed that $\gamma$-$B_{28}$ is superhard, with Vickers hardness of 50 GPa, which makes it the hardest phase of boron (the best estimates of the hardness of $\beta$-$B_{106}$ and $\alpha$-$B_{12}$ are 45 GPa [34] and 42 GPa [35], respectively). Comparison of the diffraction data on $\gamma$-$B_{28}$ with older data of Wentorf [23] shows a great deal of similarity; given also quite similar conditions of synthesis, it is very likely that what Wentorf observed was indeed $\gamma$-$B_{28}$ (in a mixture with some other phases).

---

[6]But it took much longer to publish these results. The paper, originally submitted to *Science* on December 8, 2006, was turned down and submitted to *Nature* on January 27, 2007. It took 2 years for Oganov's team to publish their manuscript in *Nature* (the paper came out on January 28, 2009). During this period, Oganov learned about Solozhenko's independent work and the two parallel teams merged.

[7]Only in compounds, such as $B_4C$, $B_6P$, $B_6O$, etc., there are roughly similar structures – but there, the two sublattices are occupied by chemically different atoms.

This discovery has attracted much attention. Very recently Zarechnaya et al. [36, 37] confirmed the structure and superhardness of $\gamma$-$B_{28}$. Their main achievement was the synthesis of micron-sized single crystals, though conditions of synthesis were suboptimal (e.g., the capsules reacted with boron sample) and their papers unfortunately contained serious errors (see [38] for details) and confusions (e.g., tetragonal T-192 and $B_{50}C_2$ are incorrectly implied to be the same phase). Their equation of state, measured to 30 GPa [36], shows large deviations from theory [1] and independent experiment [39]. Later calculations found that (*i*) the electronic spectra of the different atomic sites are indeed very different [40], confirming the charge-transfer model [1] and (*ii*) during deformation of the structure, the first bonds to break are those between the most charged atoms [41].

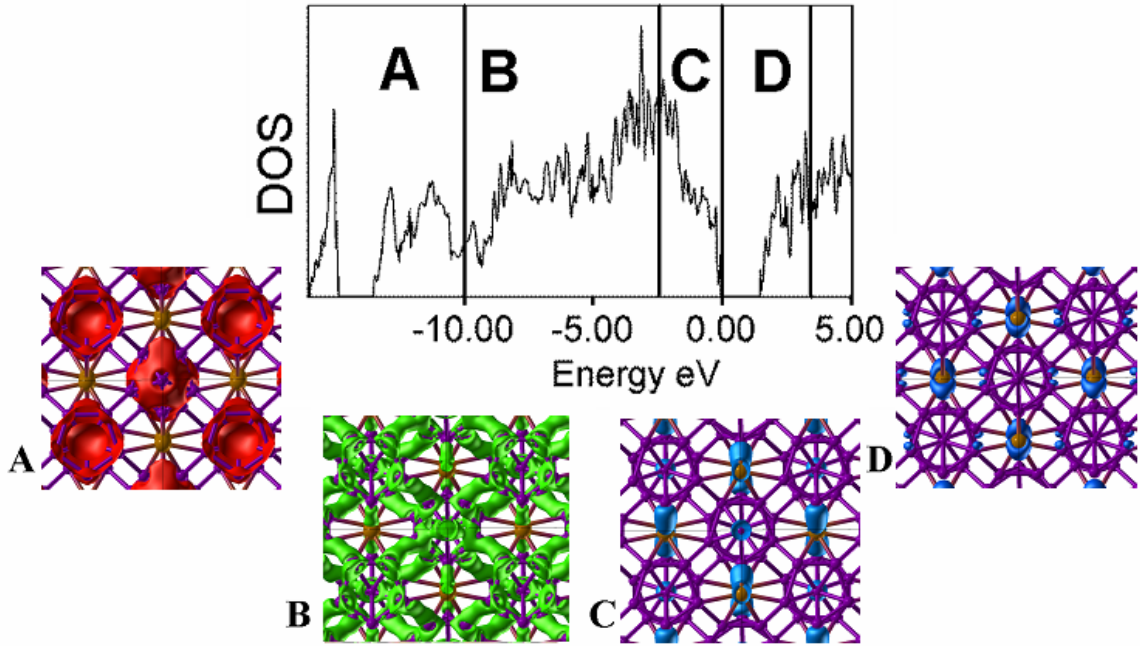

Fig. 3. Electronic structure of $\gamma$-$B_{28}$. The total density of states is shown, together with the electron density corresponding to four different energy regions denoted by letters A, B, C, D. Note that lowest-energy electrons are preferentially localized around the $B_{12}$ icosahedra, whereas highest-energy electrons are concentrated near the $B_2$ pairs.

The discovery of γ-$B_{28}$ provided the missing piece of a puzzle of the phase diagram of boron [1]. The stability field of this phase is larger than the fields of all other known boron polymorphs combined (Fig. 4). The diagram shown in Fig. 4 describes all known data in a satisfactory manner. The upper pressure limit of stability of γ-$B_{28}$ remains to be tested. Theoretical predictions of an α-Ga-type metallic phase above 74 GPa [42, 43] were confirmed [1] using crystal structure prediction tools [32], except that the predicted pressure of this phase transition was shifted to a higher value, 89 GPa, by the presence of a new phase, γ-$B_{28}$ [1]. α-Ga-type boron has been predicted to be a superconductor [44]. This structure does not contain $B_{12}$ icosahedra and heralds a new type of boron chemistry under pressure. A chemistry that remains to be probed by future experiments.

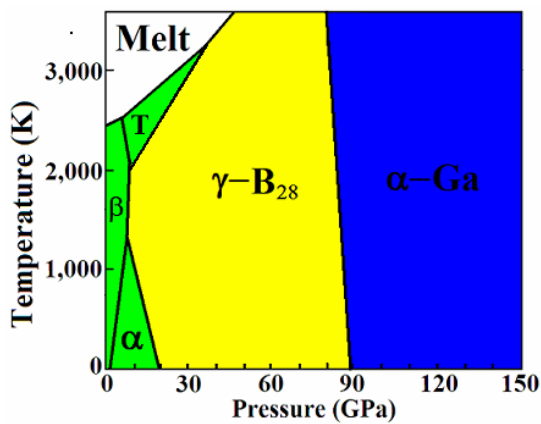

Fig. 4. Phase diagram of boron. This diagram is based on theoretical and experimental data from [1], as well as data from earlier works. Reproduced from [1].